\documentclass[twocolumn]{aastex62}

\submitjournal{The Astrophysical Journal}

\shorttitle{An Asymmetric Keplerian Disk Surrounding an O-type Protostar}
\shortauthors{Zapata et al.}

\begin{document}

\title{An Asymmetric Keplerian Disk Surrounding the O-type Protostar IRAS16547$-$4247}

\correspondingauthor{Luis A. Zapata}
\email{l.zapata@irya.unam.mx}

\author{Luis A. Zapata}
\affil{Instituto de Radioastronom\'\i a y Astrof\'\i sica, Universidad Nacional Aut\'onoma de M\'exico, P.O. Box 3-72, 58090, Morelia, Michoac\'an, M\'exico}

\author{Guido Garay}
\affiliation{Departamento de Astronom\'\i a, Universidad de Chile, Camino el Observatorio 1515, Santiago, Chile}

\author{ Aina Palau}
\affil{Instituto de Radioastronom\'\i a y Astrof\'\i sica, Universidad Nacional Aut\'onoma de M\'exico, P.O. Box 3-72, 58090, Morelia, Michoac\'an, M\'exico}

\author{Luis F. Rodr\'\i guez}
\affil{Instituto de Radioastronom\'\i a y Astrof\'\i sica, Universidad Nacional Aut\'onoma de M\'exico, P.O. Box 3-72, 58090, Morelia, Michoac\'an, M\'exico}
\affil{Mesoamerican Centre for Theoretical Physics, Universidad Aut\'onoma de Chiapas, Carretera Emiliano Zapata Km. 4 Real del Bosque, 29050 Tuxtla Guti\'errez, Chiapas, M\'exico}

\author{Manuel Fern\'andez-L\'opez}
\affiliation{Instituto Argentino de Radioastronom\'\i a (CCT-La Plata, CONICET; CICPBA), C.C. No. 5, 1894, Villa Elisa, Buenos Aires, Argentina}

\author{Robert Estalella}
\affil{Dep. de F\'{\i}sica Qu\`antica i Astrof\'{\i}sica, Institut de Ciències del Cosmos, Universitat de Barcelona, IEEC-UB, Marti i Franques 1, 08028 Barcelona, Spain}

\author{Andres Guzm\'an}
\affil{National Astronomical Observatory of Japan, National Institutes of Natural Sciences, 2-21-1 Osawa, Mitaka, Tokyo 181-8588, Japan}



\begin{abstract}

During the last decades, a great interest has emerged to know if even the most massive stars in our galaxy 
(namely the spectral O-type stars) are formed in a similar manner as the low- and intermediate-mass stars, that is,
through the presence of accreting disks and powerful outflows.  
Here, using sensitive observations of the Atacama Large Millimeter/Submillimeter Array (ALMA),
we report a resolved Keplerian disk  (with fifteen synthesized beams across its major axis) surrounding the deeply 
embedded O-type protostar IRAS16547$-$4247. The disk shows some asymmetries that could arise because of
the disk is unstable and fragmenting or maybe because of different excitation conditions within the disk.
The enclosed mass estimated from the disk Keplerian radial velocities is 25$\pm$3 M$_\odot$. 
The molecular disk is at the base of an ionized thermal radio jet and is approximately perpendicular to the jet axis orientation.   
We additionally find the existence of a binary system of compact dusty objects at the center of the accreting disk, 
which indicates the possible formation of an O-type star and a companion of lower mass.  
This is not surprising due to the high binary fraction reported in massive stars.
Subtracting the contribution of the dusty disk plus the envelope and the companion, 
we estimated a mass of 20 M$_\odot$ for the central star. 

\end{abstract}



\section{Introduction} \label{sec:intro}

\begin{figure*}
\centering
\includegraphics[angle=0, scale=0.55]{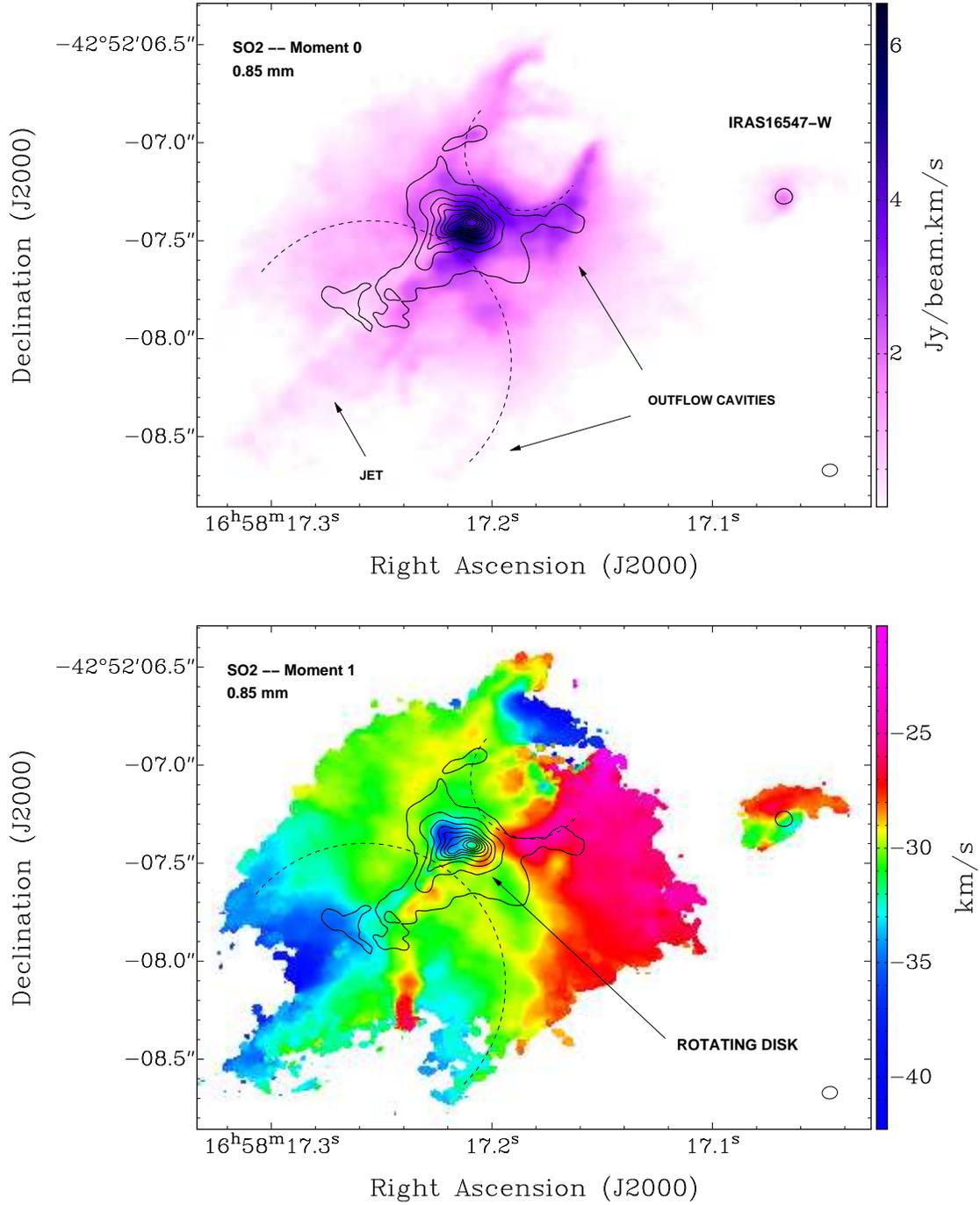} 
\caption{\scriptsize ALMA Moment 0 (upper panel) and 1 (lower panel) from the SO$_2$ v=0 (30(2,28)-29(3,27) A) thermal emission (color scale) 
overlaid with the 0.85 mm continuum emission (contours) from the object IRAS 16547$-$4247. In both panels, the contours range from 5 to 85\% 
of the peak emission, in steps of 10\%. The peak of the millimeter continuum emission is 0.58 Jy Beam$^{-1}$. The peak intensity and LSR radial 
velocity scale-bars are shown at the right. The dashed lines trace approximately the cavities created by the outflow and jet. In both panels the 
companion IRAS16547-W is shown.The half-power contour of the synthesized beam of the line image is shown in the bottom-right corner. The range 
of LSR radial velocities where we integrated to construct this image is from $-$50 to $-$15 km s$^{-1}$. } 
\end{figure*}

The first molecular structures reported to be surrounding deeply embedded O-type protostars were huge rotating toroids \citep{ces2007, bel2016, tor1983}.  
These non-equilibrium structures are large ($\sim$10,000 au) and massive (about 100 solar masses) compared
to the circumstellar disks around Solar-type stars, which have typical sizes of $\sim$100 au and masses of $\sim$0.1 M$_\odot$ \citep{luc2011}. 
The toroids are likely forming a group of stars in their centers. The detection of the toroids with these physical characteristics are a consequence of two limiting facts,
the large distances of the massive star forming regions (typically kpc) and the habitual angular resolutions (1$''$) of the first millimeter and submillimeter
interferometers ({\it e.g.} BIMA, SMA, PdBI, and CARMA). 

Recently, more sensitive and subarcsecond ALMA 
observations have revealed disk-like systems with sizes of about 1000-2000 au \citep{zap2015,jon2015,lle2016,san2018}.  Some of these disk-like structures
show rotational Keplerian-like profiles with the resulting enclosed masses corresponding to young O-type stars \citep{san2013,jon2015,lle2016,san2018}.  
However, is very important to mention that in these studies the molecular disks are only about a few synthesized beams in size, and the Keplerian 
profiles are roughly resolved.  A more dedicated study using ALMA with an angular resolution of 0.2$''$ and a high sensitivity to search for 
circumstellar, rotating disks around a small sample of six luminous ($\geq$ 10$^5$ L$_\odot$) young stellar objects
revealed a low rate of Keplerian disks around the embedded O-type stars \citep{ces2017},  suggesting that maybe 
the detection rate could be sensitive to the evolutionary stage. \citet{mau2018} proposed the presence of an SiO disk and disk wind from the O-type young star G17.64$+$0.16.
The SiO emission appears to represent a disk-like structure perpendicular to a molecular outflow traced by the $^{13}$CO.    
The position-velocity profile of the SiO is consistent with the Keplerian rotation of a disk around an object with a mass between 10-30 M$_\odot$.
A more exotic premise came from recent ALMA observations 
of W51, where the authors claim a lack of Keplerian disks, and proposed that accretion should be 
multi-directional and unsteady as modeled in hydrodynamic simulations \citep{god2018}.

IRAS16547$-$4247 is classified as a very embedded massive protostar with a total bolometric luminosity of $\sim$10$^5$ L$_\odot$ \citep{gui2003}, at a distance of 2.9 kpc.
This luminosity is equivalent to a single O8-type zero-age main-sequence star \citep{pan1973}.  Observations from single-dish telescopes revealed that the 
IRAS source coincides with a dense molecular clump with a mass of 10$^3$ M$_\odot$, contained in a radius of 0.2 pc \citep{gui2003}. More
recent observations carried out with the Submillimeter Array resolved this dense molecular clump into two compact objects with sizes of 4000 au,
IRAS16547-E and IRAS16547-W \citep{ram2009,her2014}. 
As IRAS16547-W has a very faint infrared emission in Ks band from the 2MASS catalogues and IRAS16547-E is 
the dominant source at these wavelengths (0.85 mm),
the contribution to the total luminosity of IRAS16547-W should not be important.
Further ALMA observations with an angular resolution of 0.3$''$ resolved the two millimeter continuum sources \citep{zap2015}.
The authors found that in the object IRAS16547-E, the SO$_2$ line emission, 
with low-energy transitions, E$_u$ $\leq$ 300 K,  traces the innermost parts ($\sim$ 3000 au) of a molecular NS outflow emanating 
from this region \citep{zap2015}.  At smaller scales ($\sim$1000 au) these ALMA observations also revealed a small and 
warm molecular disk-like structure that is likely driving the ionized jet and molecular outflow \citep{zap2015}. The axis of this disk-like structure is almost perpendicular
to the radio jet at the base. However, these ALMA observations failed to reveal Keplerian motions 
(that is velocities that increase as the central source is approached) in the innermost parts of the disk-like structure \citep{zap2015}. 
The extended cavities carved by the powerful radio thermal jet can also be seen in recent CH$_3$OH ALMA images \citep{hig2015} obtained with a similar spatial 
resolution than that in \citet{zap2015}. 
     
In this study, we carried out new ALMA observations with a high angular resolution at 0.85 mm ($\sim$ 0.05$''$) and a very good sensitivity to disclose the innermost 
structure of the disk candidate using high-energy transitions of Methyl Formate (CH$_3$OCHO with E$_u$ $=$ 713 K) and Carbon Monosulfide (CS with E$_u$ $=$ 1896 K).
We additionally present observations of the SO$_2$ line that show the morphology of the outflow in the vicinity of the launching zone.  

\begin{figure*}
\centering
\includegraphics[angle=0, scale=0.7]{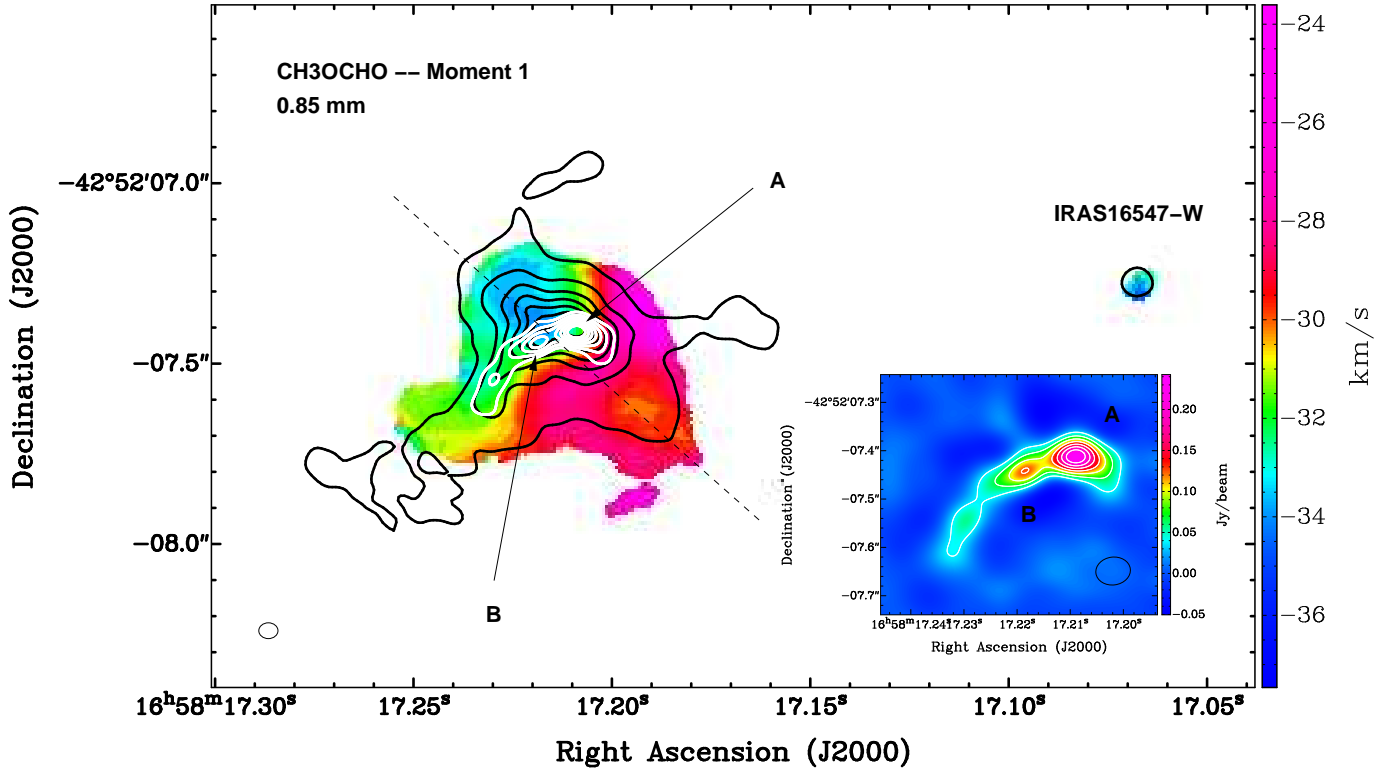} 
\caption{\scriptsize ALMA Moment 1 from the CH$_3$OCHO(37(13,25)-37(12,26) A) v=1 thermal emission (color scale) overlaid 
with the 0.85 mm continuum emission (contours) from the object IRAS 16547$-$4247.  The white contours here represent the 
continuum emission constraining the uv-range of the observations, see text. The black contours range from 5 to 85\% of the peak emission, 
in steps of 10\%.  The white contours range from 25 to 85\% of the peak emission, in steps of 10\%. The peak of the millimeter continuum 
emission is 0.25 Jy Beam$^{-1}$.  The LSR radial velocity scale-bar is shown in the right. The dashed line traces the position and orientation
 of the PV-diagrams shown in Figure 3. The half-power contour of the synthesized beam of the line image is shown in the bottom-left corner.  
 The inset in the bottom-right of the image shows the continuum image constraining the $uv$-range of the observations, as the white contours.
 The contouring is similar to the white contours in this Figure.   The half-power contour of the synthesized beam of the line image is shown in the 
 bottom-right corner. The range of LSR radial velocities where we integrated to construct this image is from $-$45 to $-$17 km s$^{-1}$.} 
\end{figure*}

\section{Observations} \label{sec:obs}

The observations of the IRAS 16547$-$4247 massive protostellar system were carried out with ALMA 
at Band 7 in 2017 August 21 and 22 as part of the Cycle 5 program 2016.1.00992.S. 
Both observations used 44 antennas with a diameter of 12 m, yielding baselines with projected lengths 
from 21 to 3696.5~m (24.7--4348.8~k$\lambda$).   The primary beam at this frequency has a full width at half-maximum (FWHM) 
of about 17.6$''$.  The continuum and molecular line emission from the massive object IRAS 16547$-$4247 fall well inside 
of the FWHM or primary beam.

The integration time on-source was about 1.6 h, and 2.5 h was used for calibration.  The continuum image was obtained averaging line-free channels
from four spectral windows (of 1.875, 0.937, 0.468, 0.468 GHz width) centered at rest frequencies: 353.684 GHz (spw0),  351.644 GHz (spw1), 
339.160 GHz (spw2), and 340.405 GHz (spw3), which cover a total bandwidth of 3.748 GHz.  However, as there are many spectral lines detected in the spectral windows,
we have likely contamination from faint lines.
These windows were chosen to observe different molecular species, such as the vibrationally excited molecule CS(7-6) v=1 at a rest frequency of 340397.96 MHz,  
SO$_2$ v=0 (30(2,28)-29(3,27) A) at a rest frequency of 340316.41 MHz, and  CH$_3$OCHO(37(13,25)-37(12,26) A) v=0 at a rest frequency of 339548.47 MHz. In this study, 
we concentrate on these spectral lines, which are outflow \citep{zha2000} and disk \citep{san2014,zap2010} tracers. 
The velocity width of each channel varied in every single spectral window:  0.430 km s$^{-1}$ for the CH$_3$OCHO, 
0.321 km s$^{-1}$ for the SO$_2$, and finally  0.215 km s$^{-1}$ for the CS.  

The weather conditions were excellent and stable for observations at band 7  with an average precipitable water vapor that varied between 0.45 and 0.5 mm 
and an average system temperature of 125 K for the first day and 140 K for the second day. The ALMA calibration included simultaneous observations of the 
183 GHz water line with water vapor radiometers, used to reduce the atmospheric phase noise. 
Quasars J1636$-$4102, J1427$-$4206, J1617-5848,  and  J1711-3744 were used to calibrate the bandpass, the atmosphere, the water vapor radiometers,
and the gain fluctuations.  J1617-5848 was also used for the flux amplitude calibration. 
The data were calibrated, imaged, and analyzed using the Common Astronomy Software Applications  CASA \citep{mac2007}.
To analyze the data, we also used the KARMA \citep{goo96} package.
Imaging of the calibrated visibilities was done using the task CLEAN.  The resulting image {\it rms} noise for the continuum was 
1.0 mJy beam$^{-1}$ at an angular 
resolution of $0\rlap.{''}06 \times 0\rlap.{''}05$ with a PA = $+$82.9$^\circ$. 
This angular resolution corresponds to a spatial scale of about 150 au at the distance of IRAS 16547$-$4247 of 2.9$\pm$0.6 kpc \citep{rod2008}. 
This is an ideal resolution to resolve circumstellar disks with a spatial size of about 1000 au \citep{zap2015}.  
   
We used the ROBUST parameter of CLEAN in CASA set to $-$2.0 in the continuum and $+0.5$ for the line images. 
For the line emission,  the resulting image {\it rms} noise was 10 mJy beam$^{-1}$  km s$^{-1}$ at an angular 
resolution of $0\rlap.{''}09 \times 0\rlap.{''}07$ with a PA = $-$76.7$^\circ$.
The resulting 0.85 mm continuum and CS, CH$_3$OCHO, and SO$_2$ line emission images are presented in Figure 1, 2, and 3.
We self-calibrated in phase and amplitude using the resulting mm continuum image as model. We then applied the solutions for self-calibration
of the continuum to the spectra data. The maximum recoverable scale for this ALMA observation is $0\rlap.{''}$87 ($\sim$ 2500 au).

\begin{figure*}
\centering
\includegraphics[angle=0, scale=0.54]{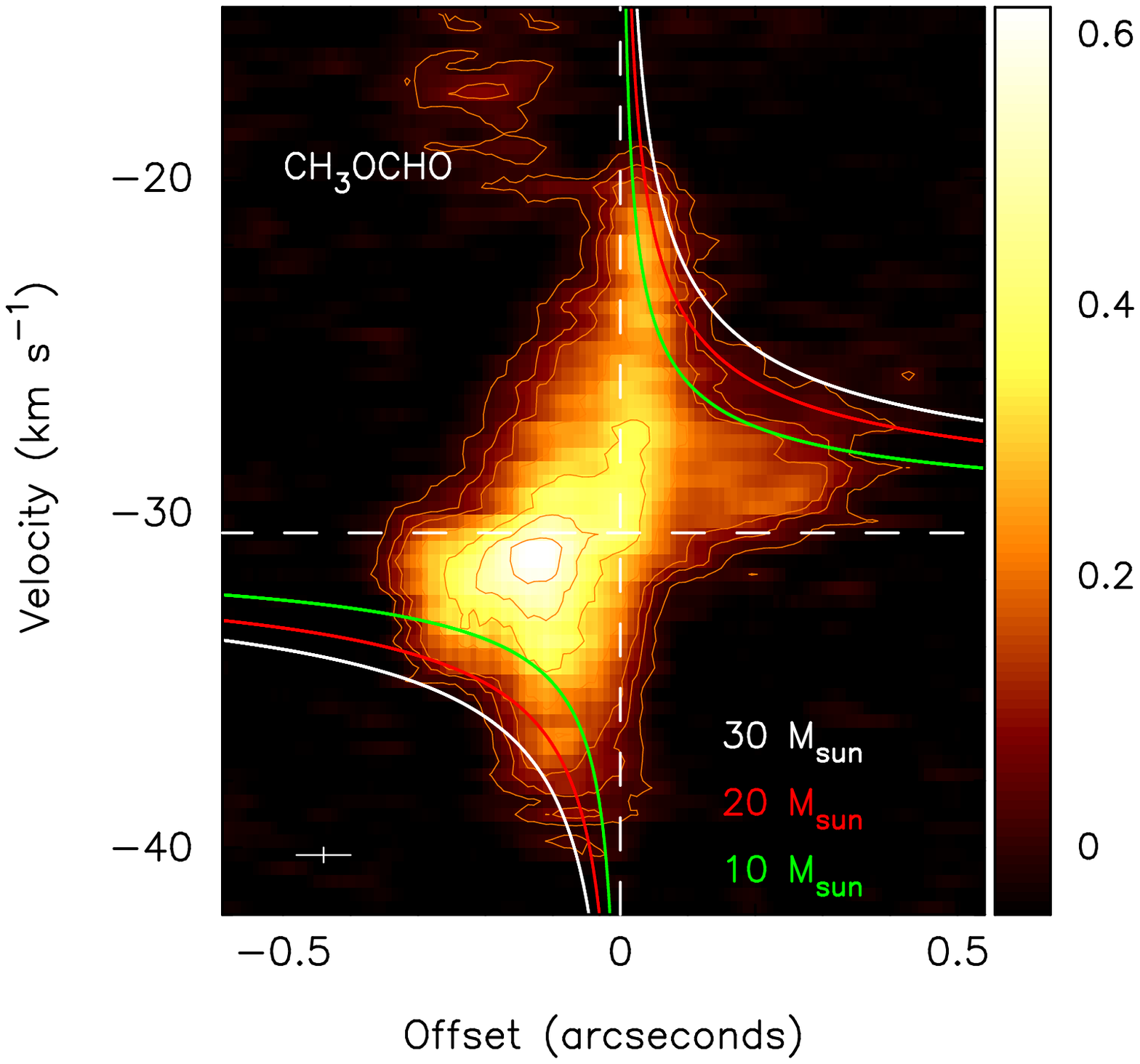} 
\includegraphics[angle=0, scale=0.53]{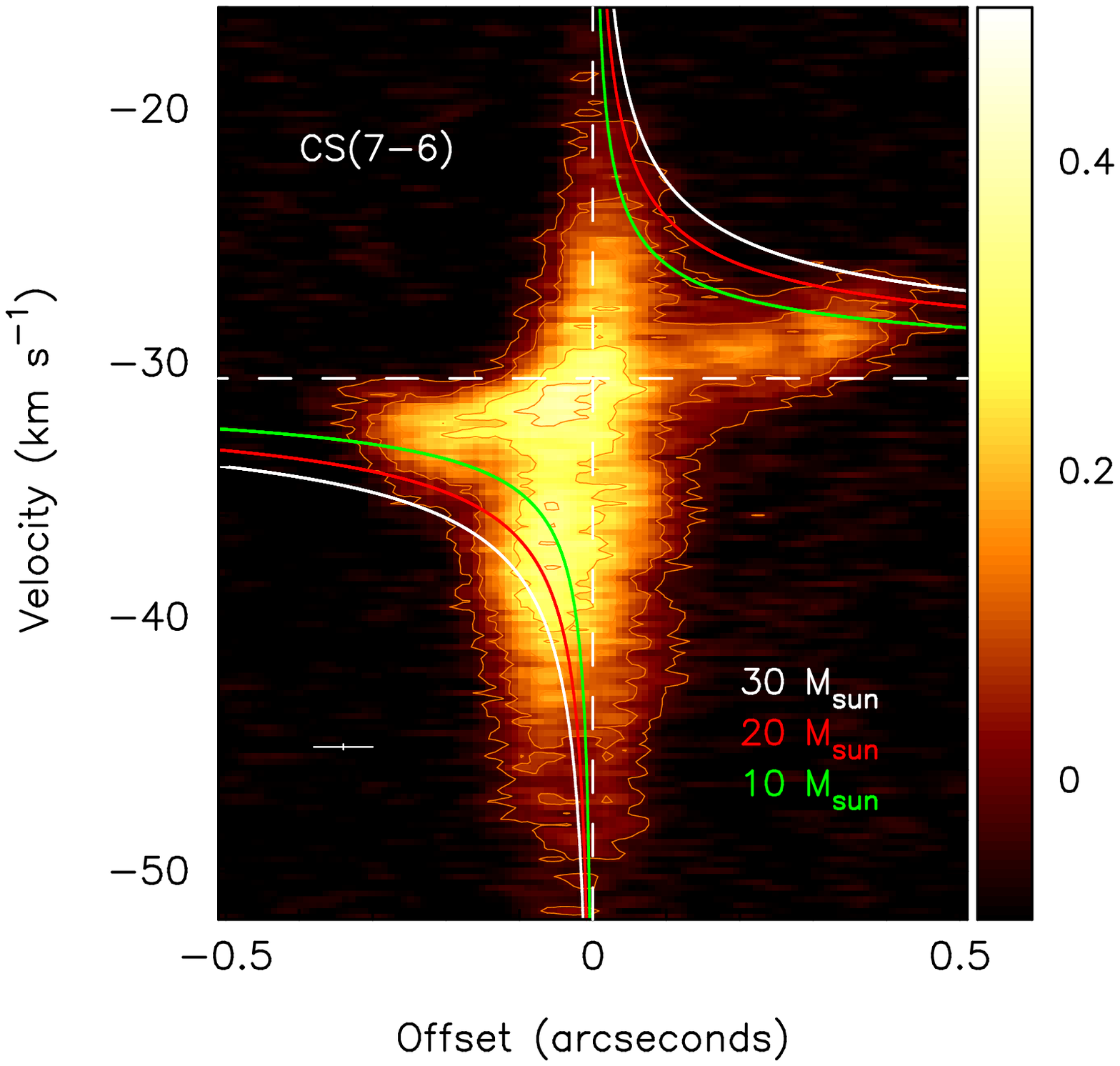}
\caption{\scriptsize ALMA Position-Velocity diagrams obtained from the CH$_3$OCHO(37(13,25)-37(12,26) A) v=0 and CS(7-6) v=1 
thermal emission and computed to a positional angle of 50$^\circ$, the mayor axis of the disk surrounding IRAS 16547-E. 
The spectral and spatial resolutions are shown in the bottom left corner. The peak intensity scale-bar is shown in the right and its units are in  
Jy Beam$^{-1}$. The white, red, and green lines mark the Keplerian rotation curves for a disk with a enclosed mass of 30 (white), 
 20 (red), and 10 (green) M$_\odot$, respectively. We assume an inclination angle for the disk of 55$^\circ$. The brown contours
  are starting from 0.05 to 0.55 mJy in steps of 0.10 mJy. The dashed lines trace the systemic velocity of the spectral lines and
  the axis center of the disk.  
 } 
\end{figure*}

\section{results and discussion} \label{sec:res}

Figure 1 shows the moment zero (integrated intensity) and one (intensity weighted velocity) maps of the SO$_2$ v=0 (30(2,28)-29(3,27) A) thermal 
emission overlaid with the 0.85 mm continuum emission as revealed by the new ALMA observations.  The 0.85 mm continuum observations traces the dust
from an extended envelope (with a spatial size of about a few 1000 au) and the disk-like structure surrounding IRAS16547-E.  The observations also 
revealed small cavities present in the northern and southern parts of its dusty envelope, likely carved by the bipolar outflow.  The bright continuum 
source is tracing the dusty disk-like structure, which is elongated in an orientation perpendicular to the orientation of the outflow at a 
Position Angle\footnote{Measured as usual counterclockwise from North to East.} (PA) of 135$^\circ$.
Additionally,  a very compact continuum object associated with IRAS16547-W is also revealed. 
The position of IRAS16547-W is at $\alpha(2000) = 16^h~ 58^m~ 17\rlap.{^s}06 \pm 0.015''; \delta(2000) = -42^\circ~ 52'~ 07\rlap.{''}285 \pm 0.001''$.
In the Appendix Section, we estimated a 
mass of the dusty envelope in IRAS16547-E of 3.55$\pm$0.05 M$_\odot$ and for IRAS16547-W (the compact object reported here) of 0.2$\pm$0.1 M$_\odot$. 

The molecular emission from SO$_2$ arising from IRAS16547-E is dramatically resolved as one can see in Figure 1. 
The dimensions ($\sim$ 10,000 au), orientation (PA$\simeq$135$^\circ$), 
and gas kinematics of this structure are consistent with those reported in Zapata et al. (2015) in the same molecule. 
However, the image presented here reveals many new features from the outflow and the disk-like structure.  Figure 1 shows the structure of the outflow
very close to IRAS16547-E with two well-defined cavities with a PA of 135$^\circ$, and in the middle of the SE cavity a collimated molecular jet is emanating 
from IRAS16547-E. All these structures are connected to the disk-like structure traced by the continuum emission.  In the very
center of the continuum source there is absorption of the SO$_2$, probably because the dust at that position is optically thick, and the molecular emission is blocked, 
as for example, in the disk of the low-mass protostar IRAS 16293$-$2224B \citep{zap2013}.  
The velocity field of SO$_2$ is also shown in Figure 1, revealing the kinematics of the outflowing molecular gas, and the disk-like structure.  
The blueshifted emission is located mostly in the SE side of the outflow, while the redshifted emission is located at the NW side. 
This image also reveals hints of outflow rotation with a similar orientation
to that of the central disk. This could be the reason why different LSR radial velocities are seen along the outflow.  For example, some redshifted emission is present in the
SW part of the outflow.    The rotating disk traced by SO$_2$ is clearly observed in the lower panel of Figure 1, right at the center of the outflow.  
It has a similar orientation (50$^\circ$), size (of about 1000 au), and gas kinematics to those values reported in Zapata et al. (2015).
However, the emission from the disk overlaps with the emission arising from the outflow and this difficults to trace its inner gas kinematics.    

Zapata et al. (2015) noted that high-energy transitions  (E$_u$ $\geq$ 500 K) seem to trace mostly
the compact disk. Therefore, we chose the molecules CH$_3$OCHO and CS to trace the circumstellar disk surrounding 
this massive embedded protostar and to study its innermost gas kinematics. Figure 2 shows the moment one
of the CH$_3$OCHO emission overlaid on the dust continuum emission presented in Figure 1.
The range of LSR radial velocities integration to construct this image is from  $-$45 to $-$17 km s$^{-1}$. The LSR systemic radial velocity of IRAS 16547-E is
$-$30.6 km s$^{-1}$ \citep{gui2003}. The image reveals the rotating disk already mapped in Zapata et al. (2015). However, from this new image, the velocity gradient
traced by the high-energy transitions reveals that there is more red- and blue-shifted velocities located close to its center.  To study the kinematics of the molecular gas traced by the 
CH$_3$OCHO and CS, we computed Position-Velocity (PV) diagrams along the major axis of this disk (PA=50$^\circ$) and they are shown in Figure 3. The PV-diagrams presented here 
probe scales of the same order as our beam size.  
These diagrams clearly revealed that the rotating disk is Keplerian with the LSR radial velocities close to its center increasing inversely proportional to the squared-root of the radius.
However, the CS 7-6 line in particular seems to show two linear features in the PV diagram (connecting the high- and low-velocity ends of the emission, respectively)  
that look like they might describe orbital rings offset from one another and not following Keplerian motions. 
One can estimate the enclosed masses from the two putative rings assuming Keplerian velocities of V$_{int}$ $=$ 10 km s$^{-1}$ at a radius r$_{int}$ $=$ 0.05$''$ (higher-velocity orbit), and   
V$_{ext}$ $=$ 2 km s$^{-1}$ at a radius r$_{ext}$ $=$ 0.3$''$ (lower-velocity orbit). We obtain M$_{int}$ $=$ 14.5 M$_\odot$ and M$_{ext}$ $=$ 3.6 M$_\odot$, respectively.  
We conclude that this interpretation must be incorrect because the enclosed mass from the exterior ring should be larger. We then suggest that the speeds are due to a single Keplerian disk. 
However, the disk shows some asymmetries that could be caused because the disk is unstable and fragmenting, as theoretically predicted for massive, large disks \citep{kra2006}
and suggested in a similar system G11.92 in \citet{lle2016, lle2018}. Furthermore, some of these asymmetries could also be caused by different excitation conditions within the disk.


In these Figures, we additionally present three Keplerian rotation curves with enclosed masses of 30,  20, and 10 M$_\odot$.  
These new probes (CH$_3$OCHO and CS) are tracing even broader LSR radial velocities ($\sim$ 40 km s$^{-1}$) within the disk than the CH$_3$SH line ($\sim$ 20 km s$^{-1}$) 
line because of the better sensitivity and angular resolution of the present observations \citep{zap2015}.  

We compute a Keplerian thin disk model fitted to the CH$_3$OCHO line data that is presented in Figure 4. We describe in detail this model in the Appendix section.  
The best fitted physical and geometrical parameters of the Keplerian disk surrounding IRAS 16547-E are presented in Table 1. With this disk model we obtained three important 
physical parameters with a relatively good accuracy:  the inclination angle 55$^\circ$ $\pm$ 5.0$^\circ$, the size 0.3$''$$\pm$0.1$''$ (1190$\pm$290 au), 
and the enclosed mass 25$\pm$3.0 M$_\odot$.  These physical values and their errors were obtained from a minimum $\chi^2$ estimation between model and data.

The twenty-five solar masses correspond to a single main-sequence O7 type star \citep{mar2005}. Subtracting the contribution of the dusty disk plus the envelope 
of about 4 M$_\odot$ (see the Appendix), and the companion \citep[0.2 to 2 M$_\odot$, for a M$_{star}$/M$_{disk}$ between 1 to 10, see:][]{rod1998,bat2018}, 
we estimated a mass of about 20 M$_\odot$ for the central star.

This is in very good agreement with the enclosed mass estimated in \citet {zap2015} with the Keplerian rotation, 
and with the rotation keplerian curves presented in Figure 3. 

 \begin{figure*}
\centering
\includegraphics[angle=0, scale=0.45]{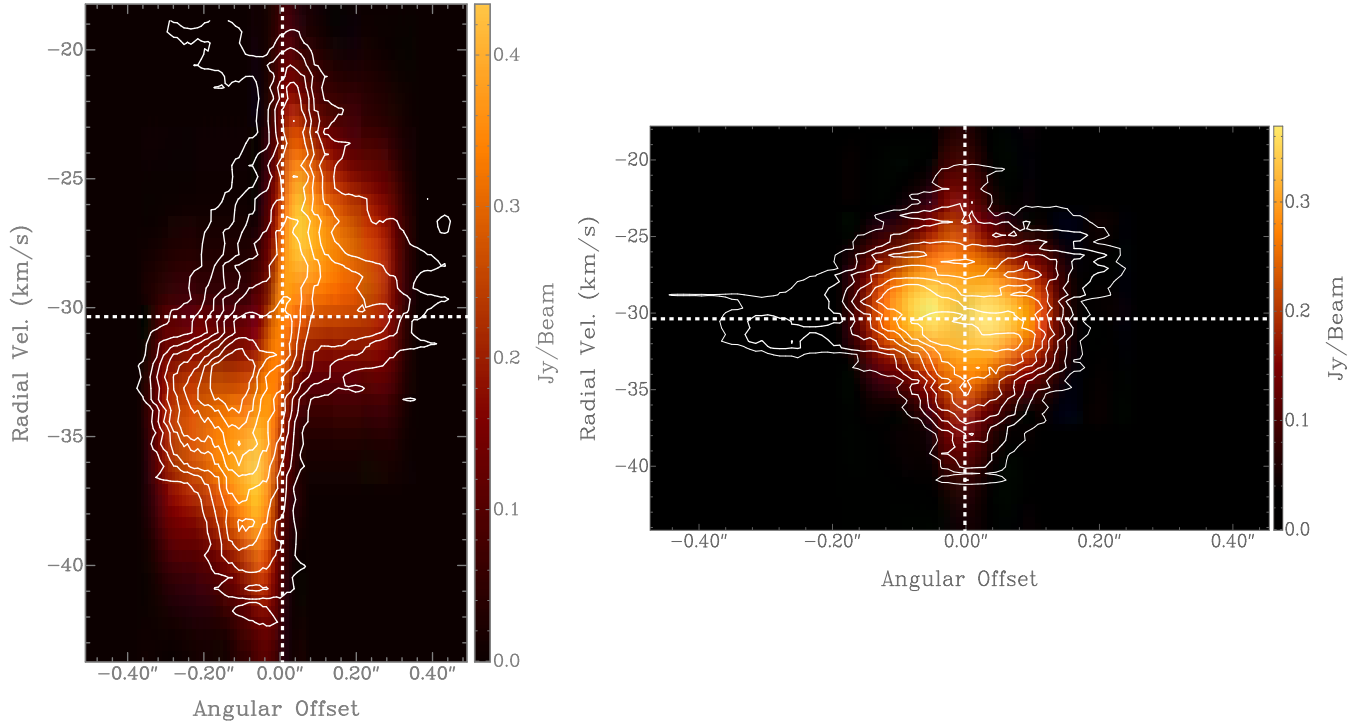} 
\caption{\scriptsize Position-Velocity diagrams obtained from the CH$_3$OCHO(37(13,25)-37(12,26) A) v=0 (contours) and the best model of the Keplerian disk (color scale). 
The PV diagrams are computed to a positional angle of 50$^\circ$ (right) and 140$^\circ$ (left). The spectral and spatial resolutions are the same as in Figure 3. 
The peak intensity scale-bar is shown in the right.  The white contours are starting from 15 to 90\% of the peak emission, in steps of 10\% for the left image, and 
from 30 to 90\% of the peak emission, in steps of 10\% for the right image.  The dashed lines trace the systemic velocity of the spectral lines and the axis center of the disk.  }
\end{figure*}  

{
In Figure 4, we compare two PV-diagrams obtained from the disk thin model and the ALMA data from the CH$_3$OCHO line. 
One PV-diagram is obtained along the disk major axis (50$^\circ$) and the other one along the perpendicular direction, i.e. the minor axis (140$^\circ$).
From these two images we find a good correspondence between our model and the data. There is however some asymmetries present in the disk 
and revealed on these images. The two most clear features are the extra-emission close to central axis and at a radial LSR velocity of about $-$33 km s$^{-1}$ and the emission close to the systemic 
velocities (see the left image of Figure 4).  These gas structures could be originated from some extra contributions as for example:  the outflow or the infalling envelope. 
This massive star is still very embedded. In the high-mass protostar IRAS 20126+4104, \citet{pal2017} showed that very complex molecules could be excited by the outflow very close to the circumstellar disk. 
These observations argue in favor of a possible contribution from the outflow in the source IRAS 16547-E to the molecular disk. This extra CH$_3$OCHO emission in the disk could be caused by
shock-driven excitation at the interface between the outflow and the envelope/disk. Given the very small scales we are tracing with our ALMA observations ($\sim$150 au), we now can reveal these new features. 

Even when the nature of the massive Source I in Orion is uncertain (this object could be a merge of protostars, see \citet{zap2009,ball2017}), it shows a Keplerian disk with a dynamical mass of 
about 7-15 M$_\odot$ \citep{hir2014,gin2018} and a comparison is warranted.  As one can see, Source I is a less massive object, probably an early B-type, however, contrary
to IRAS 16547-E its Keplearian disk is more smooth and coherent, with no obvious asymmetries.  The Keplerian disk from Source I could harbor a close binary with a separation of only 2 au \citep{ball2017}.
As for IRAS 16547-E, its molecular disk also surrounds a binary but with a larger separation and misaligned with respect to the molecular disk axis, see below. These two different geometries could
give us some hints of their formation processes.

Finally, in Figure 2, we also include a 0.85 mm continuum image of IRAS 16547$-$4247 constraining the $uv$-range (baselines $\geq$ 900 k$\lambda$). This allows
us to resolve out the most extended emission from the continuum source, and leave only structures with a scale of approximately 0.07$''$ (or 210 au).  
We find a binary source with an extended arm to its SE. We label these two sources A and B objects, with A being the brighter source at these wavelengths. 
The binary sources are separated about 0.1$''$ or 300 au in projection, IRAS 16547-A has a size of  250 au. Following again the appendix, we estimated
a mass for IRAS 16547E-A of 1 M$_\odot$ and for IRAS 16547E-B of 0.3 M$_\odot$. We thus suggest that we are tracing the inner densest part of the accreting disks detected at scales of 1000 au, 
showing that it is fragmented, with the source A likely to be related with a young O-type star, and IRAS 16547E-B associated with a lower mass object.
These young stars are still accreting and it is difficult to estimate their final masses. In the case of IRAS 16547E-A if we assume that all the 
gas and dust in the compact structure surrounding it are accreted, the mass increase will be not significant. 
The high binary fraction reported in massive stars \citep{kob2007} is in agreement with our results.

It is interesting to note that the binary orientation is orthogonal to the CH$_3$OCHO molecular disk axis, see Figure 2. 
This is contrary to what one expects from the conservation of angular momentum, where the binary should be oriented in
the same direction as the molecular disk. Similar geometries have been observed in low mass and more evolved systems, 
see \citet{lop2017}, and there is not yet a consensus of their nature.



\acknowledgments

This paper makes use of the following ALMA data:
2016.1.00992.S.  ALMA is a partnership of
ESO (representing its member states), NSF (USA) and NINS (Japan), together
with NRC (Canada), NSC and ASIAA (Taiwan), and KASI (Republic of Korea), in
cooperation with the Republic of Chile.  The Joint ALMA Observatory is
operated by ESO, AUI/NRAO and NAOJ.  LAZ acknowledge financial
support from DGAPA, UNAM, and CONACyT, M\'exico. GG acknowledge financial
support from the CONICyT Project PFB06. RE acknowledges partial financial support 
from the Spanish MINECO grant AYA2014-57369-C3 (cofunded with FEDER funds) 
and MDM-2014-0369 of ICCUB (Unidad de Excelencia "Maria de Maezru").
We are really thankful for the thoughtful suggestions of the
anonymous referee that helped to improve our manuscript.

\appendix

\begin{table}[b]
\begin{center}
{Table 1.--} Physical Parameters From The Best Keplerian Thin Disk\\ 
\bigskip
\begin{tabular}{lc}
\hline
\hline
Physical Parameter~~~~~   &  Estimated Value from the Model  \\        
\hline
Line-width ($\Delta v$)~~~~~            &  7.7$\pm$0.6 km s$^{-1}$  \\
Beam-width ($\Delta s$)~~~~~          &  0.06 arcsec (fixed)\\
Central velocity of the Disk (v$_0$)   &  $-$30.1$\pm$0.5 km s$^{-1}$\\ 
Inclination Angle ($i$)                         &  55$^\circ$ $\pm$ 5.0$^\circ$\\ 
Inner Disk Radius ($r_1$)                       &  0.005$''$(fixed)\\
Outer Disk Radius ($r_2$)                      &  0.3$''$$\pm$0.1$''$\\
Infall Velocity (v$_i$ at $r_0$)              &  $-$0.4$\pm$0.2 km s$^{-1}$\\
Rotational Velocity (v$_r$ at $r_0$)     & $+$11.6$\pm$4.0 km s$^{-1}$\\
Reference Radius ($r_0$)                    & 0.065$''$ (188 au)\\
Enclosed Mass                                   & 25 $\pm$ 3.0 M$_\odot$\\
\hline
\hline
\end{tabular}
\end{center}
\end{table}

\section{Dust Mass Estimation}

Assuming that the dust emission is optically thin and isothermal, 
the dust mass (M$_d$) is directly proportional to the flux density (S$_\nu$) as:

$$
M_d=\frac{D^2 S_\nu}{\kappa_\nu B_\nu(T_d)},
$$    

\noindent
where $D$ is the distance to the object, $\kappa_\nu$ is the dust
mass opacity, and B$_\nu(T_d)$ is the Planck function for the dust
temperature T$_d$.  Assuming a dust mass opacity ($\kappa_\nu$) of 0.020
cm$^2$ g$^{-1}$ (taking a gas-to-dust ratio of 100) appropriate  for these wavelengths (0.85 mm) 
for coagulated dust particles \citep{Oss1994}, 
as well as a characteristic dust temperature (T$_d$) of 250 K \citep{her2014},  and for flux densities 
estimated from a Gaussian fit for IRAS16547-E of 1.05 Jy, and for IRAS16547-W of 0.020 Jy,
we estimated a lower limit (because the emission is probably not optically thin) for the mass of the 
most compact part (i.e., not interferometrically filtered) of the disk and envelope system associated with IRAS16547-E 
of about  3.55$\pm$0.05 M$_\odot$ and of 0.2$\pm$0.5 M$_\odot$ for the most 
compact object associated with IRAS16547-W. The dust mass estimated here comes from several synthesized beams.
The spectral index was estimated using the spw0 and spw3, with central frequencies of 353.684 and 340.405 GHz, respectively.  

Following the radio observations presented in \citet{rod2005}, we can estimate the expected radio emission associated with IRAS 16547-W
at the submillimeter wavelengths. \citet{rod2005} reported that this source has a spectral index of 0.33, which they suggested to be related with 
a thermal jet. Hence, we estimate a density flux of 30 mJy for these wavelengths, a very small contribution ($\sim$ 3\%) at these wavelengths.
We then conclude that most of the emission arise from thermal dust.

Here, with our spatial resolution of about 150 au,
we resolved out part of the continuum emission, as it can be seen from recent ALMA observations, see \citet{zap2015}. 
Note that the level of uncertainty in the given lower limits is only statistical. The actual uncertainty could be very large (a factor of up to 4) 
given the uncertainty in the opacity and in the dust temperature.  


\section{Keplerian Thin Disk Model}

We have used the model of a thin disk with a Keplerian and infall kinematics made by Robert 
Estalella\footnote{http://www.am.ub.edu/~robert/Documents/thindisk.pdf}. 
This model considers a geometrically thin disk with an inner radius $r_1$ and an outer radius $\it r_2$. 
The inclination angle of the disk is defined as the angle (i) formed by the line-of-sight and the disk axis ($i=0\deg$ for a face-on disk, $i=90\deg$ for an edge-on disk). 
In this model two motions are considered:  the first  
is the rotation velocity of the disk, which here, we assume is Keplerian and given by a 
power law of the radius, $v_r(r/r_0)^{-0.5}$, where $r_0$ is the reference radius and $v_r$ is the rotation velocity at such radius; 
the second one is the infall toward the disk center, with an infall velocity also given by a power law, $v_i(r/r_0)^{-0.5}$, where
$v_i$ is the infall velocity at the reference radius.              

The kinematics of the disk is computed by considering a cube of
velocity channels and, for each channel, a grid of points onto the plane of the
disk. For each point of the grid, the projection of its rotation and infall
velocities onto the line of sight, $v_z$, is calculated. A Gaussian line profile
of width $\Delta v$, centered on $v_z$, and with an intensity proportional to 
$(r/r_0)^{q_d}$,  is added to the channels associated
with the grid point. After the cube onto the plane of the disk is computed, its
projection onto a cube with a grid onto the plane of the sky is estimated.
Finally, each channel of the plane-of-the-sky cube is convolved with a Gaussian
beam of width $\Delta s$.
The intensity scale depends on a factor scale, which is estimated by
minimizing the sum of the squared differences between the data cube and the
model cube.

The disk model depends on 13 parameters, however, here we assumed 5 of them, the
position of the disk is at $(0,0)$, the power-law indices of the rotation and
infall velocities are equal to $0.5$, and the power-law index of the intensity
is $1.0$. And some others can be guessed beforehand, such as $\Delta s$ and
$\Delta v$. Using a similar procedure as that described in \citet{est2012} and \citet{san2013b}, 
the $m$-dimensional parameter space is
searched for the minimum of the rms fit residual for all the channel maps. Once
a minimum of the rms fit residual is found, the uncertainty in the parameters
fitted is found as the increment of each of the  parameters of the fit necessary
to increase the rms fit residual a factor of $[1+\Delta(m,\alpha)/(n-m)]^{1/2}$ 
\citep{wall2003,san2013b}, where $n$ is the number of
data points fitted, $m$ is the number of parameters fitted, and  $\Delta(m,
\alpha)$ is the value of $\chi^2$ for $m$ degrees of freedom  (the number of
free parameters) and $\alpha$ is the significance level, $\alpha=0.68$
(equivalent to $1\sigma$ for a Gaussian error distribution).

In Table 1, we give the best fit values and 
their statistical errors. In Figure 4, we show the PV-diagrams (computed at the same PA as in Figure 3 and perpendicular) of the best fitted model, 
where it is revealed that the model and the data agree reasonably well. However, we note that there could be an extended infalling envelope 
surrounding the Keplerian disk, which could be producing the redshifted emission close to the disk center, see Figure 4, which is 
not present in our disk model.      

As described in \citet{gir2014}, the model assumes that the intensity of the disk is directly proportional to the  
depth along the line of sight, or in other words the emission is optically thin, and that the density and temperature are described by 
a single temperature and density, respectively. At these wavelengths, the CH$_3$OCHO and CS emission is most likely to be optically thick 
in the very center of the disk, so the inner disk radius found in the model could be due only to opacity effects.

\end{document}